\documentclass[10pt,a4paper]{article}

\usepackage[latin1]{inputenc} 
\usepackage[dvips]{graphicx} 
\usepackage{a4wide}
\usepackage{latexsym}
\usepackage{amsmath}
\usepackage{amssymb}

\title{Tensor charges of light baryons in the Infinite Momentum Frame}

\author{Cédric Lorcé\\ \small{\emph{Université de
Liège, Institut de Physique, Bât. B5a, B4000 Liège, Belgium}}\\
\small{\emph{Ruhr-Universität Bochum, Institut für Theoretische Physik II, D-44780 Bochum, Germany}}\\
\small{\emph{E-mail: C.Lorce@ulg.ac.be}}}
\date{}

\newcommand{\ud}{\mathrm{d}}
\newcommand{\uL}{\mathcal{L}}
\newcommand{\uM}{\mathcal{M}}

\newcommand{\uQcal}{\mathcal{Q}}
\newcommand{\uN}{\mathcal{N}}

\newcommand{\pslash}{p\!\!\!/}

\newcommand{\usigma}{\boldsymbol{\sigma}}

\newcommand{\ux}{\mathbf{x}}

\newcommand{\un}{\mathbf{n}}
\newcommand{\up}{\mathbf{p}}

\newcommand{\uq}{\mathbf{q}}
\newcommand{\uQ}{\mathbf{Q}}

\begin{document}

\maketitle

\begin{center}
\begin{minipage}[t]{15cm}
\small{We have used the Chiral-Quark Soliton Model formulated in the
Infinite Momentum Frame to investigate the octet, decuplet and
antidecuplet tensor charges up to the $5Q$ level. Using flavor
$SU(3)$ symmetry we have obtained for the proton $\delta u=1.172$
and $\delta d=-0.315$ in fair agreement previous model estimations.
The $5Q$ allowed us to estimate also the strange contribution to the
proton tensor charge $\delta s=-0.011$. All those values have been
obtained at the model scale $Q^2_0=0.36$ GeV$^2$.}
\end{minipage}
\end{center}

\section{Introduction}

Nucleon properties are characterized by its parton distributions in
hard processes. At the leading twist level there have been
considerable efforts both theoretically and experimentally to
determine the unpolarized $f_1(x)$ and longitudinally polarized (or
helicity) $g_1(x)$ quark-spin distributions. In fact a third
structure function exists and is called the transversity
distribution $h_1(x)$ \cite{Ralston}. The functions $f_1,g_1,h_1$
are respectively spin-average, chiral-even and chiral-odd spin
distributions. Only $f_1$ and $g_1$ contribute to deep-inelastic
scattering (DIS) when small quark-mass effects are ignored. The
function $h_1$ can be measured in certain physical processes such as
polarized Drell-Yan processes \cite{Ralston} and other exclusive
hard reactions \cite{Jaffe,Hard,Collins}. Let us stress however that
$h_1(x)$ does not represent the quark transverse spin distribution.
The transverse spin operator does not commute with the free-particle
Hamiltonian. In the light-cone formalism the transverse spin
operator is a bad operator and depends on the dynamics. This would
explain why the interest in transversity distributions is rather
recent. The interested reader can find a review of the subject in
\cite{Review}.

The present study was performed in the framework of Chiral-Quark
Soliton Model ($\chi$QSM) where a baryon is seen as three
constituent quarks bound by a self-consistent mean classical pion
field \cite{Profile function}. It is fully relativistic and
describes in a natural way the quark-antiquark sea. This model has
been recently formulated in the infinite momentum frame (IMF)
\cite{PetPol,DiaPet}. This provides a new approach for extracting
pre- and postdictions out of the model. The infinite momentum frame
formulation is attractive in many ways. For example light-cone wave
functions are particularly well suited to compute matrix elements of
operators. One can even choose to work in a specific frame where the
annoying part of currents, \emph{i.e.} pair creation and
annihilation part, does not contribute. On the top of that it is in
principle also easy to compute parton distributions once light-cone
wave functions are known. The technique has already been used to
study vector and axial charges of the nucleon and $\Theta^+$
pentaquark width up to the $7Q$ component \cite{DiaPet,Moi,Moi3}. It
has been shown that relativistic effects (\emph{i.e.} quark angular
momentum and additional quark-antiquark pairs) are non-negligible.
For example they explain the reduction of the naïve quark model
value 5/3 for the nucleon axial charge $g^{(3)}_A$ down to a value
close to $1.257$ observed in $\beta$ decays.

In this paper we present our results concerning octet, decuplet and
antidecuplet tensor charges. We briefly explain the $\chi$QSM
approach on the light cone and give explicit definition of
quantities needed for the computation in Section \ref{Section deux}.
Then in Section \ref{Section trois} we discuss a little bit tensor
charges and remind Soffer's inequality. We proceed in Section
\ref{Section quatre} with a discussion on Melosh rotation usually
used in light-cone models compared to the $\chi$QSM where angular
momentum with dynamical origin is naturally encoded. In Sections
\ref{Section cinq} and \ref{Section six} we explain how we can
compute matrix elements and express the physical quantities as
linear combination of a few scalar overlap integrals. Our final
results can be found in Section \ref{Section sept} where they are
compared with the sole experimental extraction achieved up to now.

\section{$\chi$QSM on the Light Cone}\label{Section deux}

Chiral-Quark Soliton Model ($\chi$QSM) is a model proposed to mimic
low-energy QCD. It emphasizes the role of constituent quarks of mass
$M$ and pseudoscalars mesons as the relevant degrees of freedom and
is based on the following effective Lagrangian
\begin{equation}
\uL_{\chi QSM}=\bar\psi(p)(\pslash-MU^{\gamma_5})\psi(p)
\end{equation}
where $U^{\gamma_5}$ is a $SU(3)$ matrix
\begin{equation}
U^{\gamma_5}=\left(\begin{array}{cc}
                     U_0 & 0 \\
                     0 & 1
                   \end{array}
\right),\qquad U_0=e^{i\pi^a\tau^a\gamma_5}
\end{equation}
and $\tau^a$ are the usual $SU(2)$ Pauli matrices. In this model
constituent quarks are bound by a relativistic mean pion field
$U^{\gamma_5}$ that has a non-trivial topology, \emph{i.e.} the pion
field is a soliton.

Within this model it has been shown \cite{PetPol,DiaPet} that one
can write a general expression for $SU(3)$ baryon wave functions
\begin{equation}\label{Wavefunction}
|\Psi_B\rangle=\prod_{\textrm{color}=1}^{N_C}\int(\ud\up)F(\up)a^\dag(\up)\exp\left(\int(\ud\up)(\ud\up')\,a^\dag(\up)W(\up,\up')b^\dag(\up')\right)|\Omega_0\rangle.
\end{equation}
This expression may look somewhat complicated at first view but in
fact it is really transparent. The model describes baryons as $N_C$
quarks populating the valence level whose wave function is $F$
accompanied by a whole sea of quark-antiquarks represented by the
exponential. The wave function of such a quark-antiquark pair is
$W$. We intentionally did not put the spin, isospin, flavor and
color indices to keep things simple. The full expression can be
found in \cite{DiaPet}. This wave function is supposed to encode a
lot of information about all light baryons.

\subsection{Valence wave function}
\begin{figure}[h]\begin{center}\begin{minipage}[c]{8cm}\begin{center}\includegraphics[width=8cm]{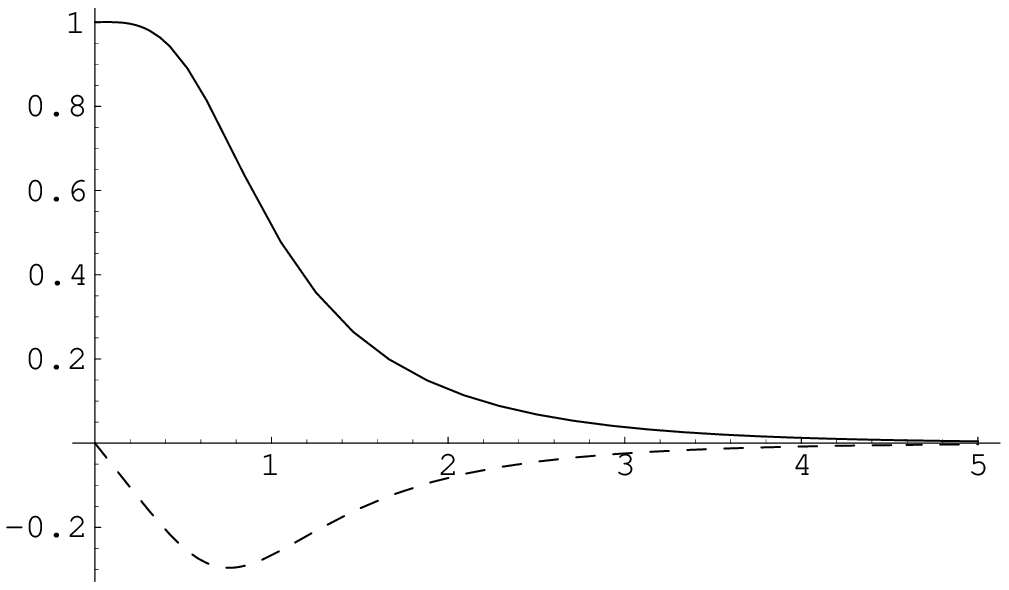}\caption{\small{Upper $s$-wave component $h(r)$ (solid) and lower $p$-wave component
$j(r)$ (dashed) of the bound-state quark level in light baryons.
Each of the three valence quarks has energy $E_\textrm{lev}=200$
MeV. Horizontal axis has units of $1/M=0.57$ fm.}}\label{Level}
\end{center}\end{minipage}\hspace{0.5cm}
\begin{minipage}[c]{8cm}\begin{center}\includegraphics[width=8cm]{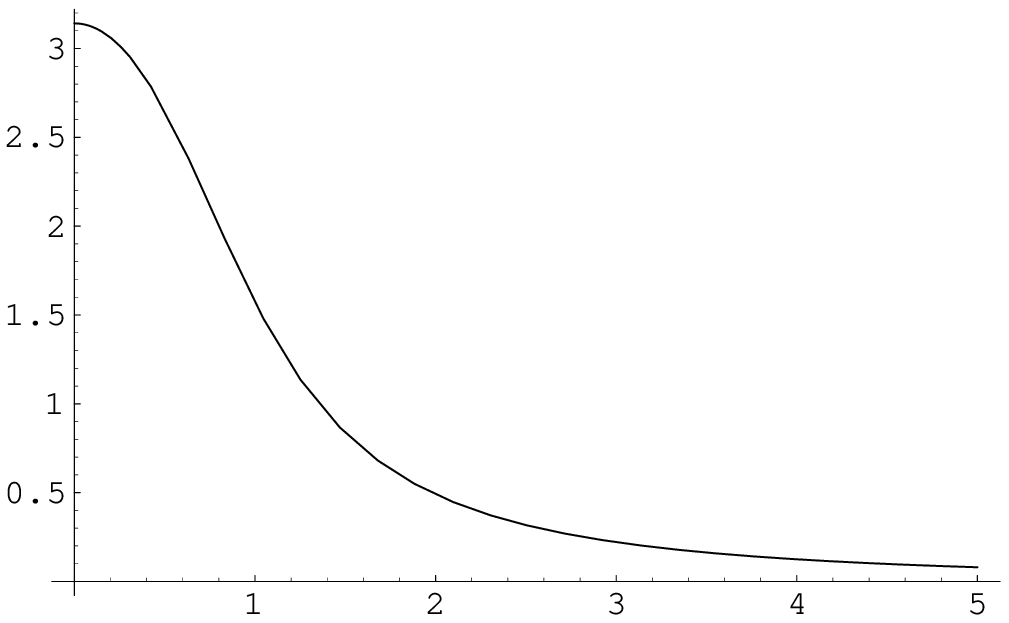}\caption{\small{Profile of
the self-consistent chiral field $P(r)$ in light baryons. The
horizontal axis unit is $r_0=0.8/M=0.46$
fm.\newline\newline\newline}}\label{Profile}
\end{center}\end{minipage}\end{center}
\end{figure}

On the light cone the valence level wave function $F$ is given by
\begin{equation}\label{Discrete level IMF}
F^{j\sigma}_\textrm{lev}(z,\up_\perp)=\sqrt{\frac{\uM}{2\pi}}\left[\epsilon^{j\sigma}h(p)+(p_z\bold
1+i\up_\perp\times\tau_\perp)^\sigma_{\sigma'}\epsilon^{j\sigma'}\frac{j(p)}{|\up|}\right]_{p_z=z\uM-E_\textrm{lev}}
\end{equation}
where $j$ and $\sigma$ are respectively isospin and spin indices,
$z$ is the fraction of baryon longitudinal momentum carried by the
quark, $\up_\perp$ is the transverse momentum and $\uM$ is the
classical soliton mass. The functions $h(p)$ and $j(p)$ are Fourier
transforms of the upper ($L=0$) $h(r)$ and lower ($L=1$) $j(r)$
component of the spinor solution (see Fig.\ref{Level}) of the static
Dirac equation with eigenenergy\footnote{This eigenenergy turned out
to be $\approx 200$ MeV when solving the system of equations
self-consistently.} $E_\textrm{lev}$
\begin{equation}
\psi_\textrm{lev}(\ux)=\left(\begin{array}{c}\epsilon^{ji}h(r)\\-i\epsilon^{jk}(\un\cdot\sigma)^i_k\,
j(r)\end{array}\right),\qquad\left\{\begin{array}{c}h'+h\,M\sin
P-j(M\cos P+E_\textrm{lev})=0\\j'+2j/r-j\,M\sin P-h(M\cos
P-E_\textrm{lev})=0\end{array}\right.
\end{equation}
where $P(r)$ is the profile function of the soliton
\begin{equation}\label{Self-consistent field}
\pi(\ux)=\pi^a(\ux)\tau^a=\un\cdot\tau\,P(r),\qquad\un=\ux/r,\qquad
r=\sqrt{\ux^2}.
\end{equation}
This profile function is fairly approximated by \cite{Profile
function, Approximation} (see Fig.\ref{Profile})
\begin{equation}\label{Profile function}
P(r)=2\arctan\left(\frac{r_0^2}{r^2}\right),\qquad
r_0\approx\frac{0.8}{M}.
\end{equation}

\subsection{Pair wave function}
The quark-antiquark pair wave function $W$ can be written in terms
of the Fourier transform of the pion field with chiral circle
condition $\Pi^2+\Sigma^2=1$, $U_0=\Sigma+i\Pi\gamma_5$. The pion
field is then given by
\begin{equation} \Pi=\un\cdot\tau\sin P(r),\qquad \Sigma(r)=\cos P(r)
\end{equation}
and its Fourier transform by
\begin{equation}\label{Fourier tranform of mean field}
\Pi(\uq)^j_{j'}=\int\ud^3\ux\,
e^{-i\uq\cdot\ux}(\un\cdot\tau)^j_{j'}\sin
P(r),\qquad\Sigma(\uq)^j_{j'}=\int\ud^3\ux\, e^{-i\uq\cdot\ux}(\cos
P(r)-1)\delta^j_{j'}
\end{equation}
where $j$ and $j'$ are the isospin indices of the quark and
antiquark respectively. The pair wave function appears as a function
of the fractions of the baryon longitudinal momentum carried by the
quark $z$ and antiquark $z'$ of the pair and their transverse
momenta $\up_\perp$, $\up'_\perp$
\begin{equation}
W^{j,\sigma}_{j'\sigma'}(z,\up_\perp;z',\up'_\perp)=\frac{M\uM}{2\pi
Z}\left\{\Sigma^j_{j'}(\uq)[M(z'-z)\tau_3+\uQ_\perp\cdot\tau_\perp]^\sigma_{\sigma'}+i\Pi^j_{j'}(\uq)[-M(z'+z)\bold
1+i\uQ_\perp\times\tau_\perp]^\sigma_{\sigma'}\right\}
\end{equation}
where $\uq=((\up+\up')_\perp,(z+z')\uM)$ is the three-momentum of
the pair as a whole transferred from the background fields
$\Sigma(\uq)$ and $\Pi(\uq)$. As earlier $j$ and $j'$ are isospin
and $\sigma$ and $\sigma'$ are spin indices with the prime for the
antiquark. In order to condense the notations we used
\begin{equation}\label{Notation}
Z=\uM^2zz'(z+z')+z(p'^2_\perp+M^2)+z'(p^2_\perp+M^2),\qquad
\uQ_\perp=z\up'_\perp-z'\up_\perp.
\end{equation}
A more compact form for this wave function can be obtained by means
of the following two variables
\begin{equation}
y=\frac{z'}{z+z'},\qquad\uQcal_\perp=\frac{z\up'_\perp-z'\up_\perp}{z+z'}.
\end{equation}
The pair wave function then takes the form
\begin{equation}\label{Pair wavefunction}
W^{j,\sigma}_{j'\sigma'}(y,\uq,\uQcal_\perp)=\frac{M\uM}{2\pi
}\frac{\Sigma^j_{j'}(\uq)[M(2y-1)\tau_3+\uQcal_\perp\cdot\tau_\perp]^\sigma_{\sigma'}+i\Pi^j_{j'}(\uq)[-M\bold
1+i\uQcal_\perp\times\tau_\perp]^\sigma_{\sigma'}}{\uQcal^2_\perp+M^2+y(1-y)\uq^2}.
\end{equation}

\subsection{Rotational wave function}
To obtain the wave function of a specific baryon with given spin
projection, one has to rotate the soliton in ordinary and flavor
spaces and then project on quantum numbers of this specific baryon.
For example, one has to compute the following integral to obtain the
neutron rotational wave function in the $3Q$ sector
\begin{equation}
T(n^0)_{k,j_1j_2j_3}^{f_1f_2f_3}=\int\ud R\,n_k(R)^*
R^{f_1}_{j_1}R^{f_2}_{j_2}R^{f_3}_{j_3}
\end{equation}
where $R$ is a $SU(3)$ matrix and
$n_k(R)^*=\frac{\sqrt{8}}{24}\,\epsilon_{kl}R^{\dag l}_2R^3_3$
represents the way that the neutron is transformed under $SU(3)$
rotations. This integral means that the neutron state $n_k(R)^*$ is
projected on the $3Q$ sector
$R^{f_1}_{j_1}R^{f_2}_{j_2}R^{f_3}_{j_3}$ by means of the
integration over all $SU(3)$ matrices $\int\ud R$. By contracting
this rotational wave function $T(n^0)_{k,j_1j_2j_3}^{f_1f_2f_3}$
with the nonrelativistic $3Q$ wave function
$\epsilon^{j_1\sigma_1}\epsilon^{j_2\sigma_2}\epsilon^{j_3\sigma_3}h(p_1)h(p_2)h(p_3)$
one finally obtains the non relativistic neutron wave function
\begin{equation}
|n^0\rangle_k^{f_1f_2f_3,\sigma_1\sigma_2\sigma_3}=\frac{\sqrt{8}}{24}\,\epsilon^{f_1f_2}\epsilon^{\sigma_1\sigma_2}\delta^{f_3}_2\delta^{\sigma_3}_kh(p_1)h(p_2)h(p_3)+\textrm{cyclic
permutations of 1,2,3.}
\end{equation}
This expression means\footnote{One has $f=u,d,s$ and
$\sigma=\uparrow,\downarrow$.} that there is a $ud$ pair in
spin-isospin zero combination
$\epsilon^{f_1f_2}\epsilon^{\sigma_1\sigma_2}$ and that the third
quark is a down quark $\delta^{f_3}_2$ and carries the whole spin of
the neutron $\delta^{\sigma_3}_k$. This is in fact exactly the
$SU(6)$ spin-flavor wave function for the neutron.

In the $5Q$ sector the neutron wave function in the momentum space
is given by
\begin{eqnarray}
(|n\rangle_k)^{f_1f_2f_3f_4,\sigma_1\sigma_2\sigma_3\sigma_4}_{f_5,\sigma_5}(\up_1\ldots\up_5)&=&\frac{\sqrt{8}}{360}\,F^{j_1\sigma_1}(\up_1)F^{j_2\sigma_2}(\up_2)F^{j_3\sigma_3}(\up_3)W^{j_4\sigma_4}_{j_5\sigma_5}(\up_4,\up_5)\nonumber\\
&\times&\epsilon_{k'k}\left\{\epsilon^{f_1f_2}\epsilon_{j_1j_2}\left[\delta^{f_3}_2\delta^{f_4}_{f_5}\left(4\delta^{j_5}_{j_4}\delta^{k'}_{j_3}-\delta^{j_5}_{j_3}\delta^{k'}_{j_4}\right)+\delta^{f_4}_2\delta^{f_3}_{f_5}\left(4\delta^{j_5}_{j_3}\delta^{k'}_{j_4}-\delta^{j_5}_{j_4}\delta^{k'}_{j_3}\right)\right]\right.\nonumber\\
&+&\left.\epsilon^{f_1f_4}\epsilon_{j_1j_4}\left[\delta^{f_2}_2\delta^{f_3}_{f_5}\left(4\delta^{j_5}_{j_3}\delta^{k'}_{j_2}-\delta^{j_5}_{j_2}\delta^{k'}_{j_3}\right)+\delta^{f_3}_2\delta^{f_2}_{f_5}\left(4\delta^{j_5}_{j_2}\delta^{k'}_{j_3}-\delta^{j_5}_{j_3}\delta^{k'}_{j_2}\right)\right]\right\}\nonumber\\
&+&\textrm{permutations of 1,2,3}.
\end{eqnarray}
The color degrees of freedom are not explicitly written but the
three valence quarks (1,2,3) are still antisymmetric in color while
the quark-antiquark pair (4,5) forms a color singlet. Let us
concentrate on the flavor part of this wave function. One can notice
that it allows hidden flavors to access to the valence level. The
flavor structure of the neutron at the $5Q$ level is
\begin{equation}\label{hidden}
|n\rangle=A|udd(u\bar u)\rangle+B|udd(d\bar d)\rangle+C|udd(s\bar
s)\rangle+D|uud(d\bar u)\rangle+E|uds(d\bar s)\rangle
\end{equation}
where the three first flavors belong to the valence sector and the
last two to the quark-antiquark pair. All rotational wave functions
up to the $7Q$ sector can be found in the Appendix of \cite{Moi3}.

\section{Tensor charge and Soffer's Inequality}\label{Section trois}

Let us consider a nucleon travelling in the $z$ direction with its
polarization in the $x$ direction. One can classify the quark
polarizations in terms of the transversity eigenstates
$|\uparrow\rangle=(|+\rangle+|-\rangle)/\sqrt{2}$ and
$|\downarrow\rangle=(|+\rangle-|-\rangle)/\sqrt{2}$ where
$|+\rangle$ and $|-\rangle$ are the usual helicity eigenstates. One
defines the axial and tensor charges as the first moment of helicity
and transversity distributions
\begin{eqnarray}
\Delta q&=&\int^1_0\ud x\,[g_1(x)+\bar g_1(x)]=\int^1_0\ud
x\,[N_+(x)-N_-(x)+\bar N_+(x)-\bar N_-(x)],\\
\delta q&=&\int^1_0\ud x\,[h_1(x)-\bar h_1(x)]=\int^1_0\ud
x\,[N_\uparrow(x)-N_\downarrow(x)-\bar N_\uparrow(x)+\bar
N_\downarrow(x)]
\end{eqnarray}
where $N_{\uparrow,\downarrow,+,-}(x)$ is the density of quarks with
polarization $|\uparrow,\downarrow,+,-\rangle$. The quantity $\delta
q$ then counts valence quarks of opposite transversity. The sea
quarks do not contribute because the quark tensor operator $\bar\psi
i\sigma^{\mu\nu}\gamma^5\psi$ is odd under charge conjugation. This
has to be contrasted with $\Delta q$ whose quark axial operator
$\bar\psi\gamma^\mu\gamma^5\psi$ is chiral even and thus includes
the sea polarization. One can then write $\Delta q=\Delta
q_{val}+\Delta q_{sea}$ and $\delta q=\delta q_{val}$.

In the nonrelativistic Naive Quark Model (NQM) one has the identity
$\delta q_{NR}=\Delta q_{NR}$ because of rotational invariance.
However relativistic effects break this invariance and introduce a
difference between the actual charges $\Delta q_{val}\neq\delta q$
and $\Delta q_{val}\neq\Delta q_{NR}$.

There are several theoretical determinations using the MIT bag model
\cite{Jaffe,Hard}, QCD sum rules \cite{Ioffe}, a chiral
chromodielectric model \cite{Barone}, the $\chi$QSM \cite{Kim}, on
the light cone by means of the Melosh rotation \cite{Schmidt}, using
axial vector mesons \cite{Gamberg} or in a quark-diquark model
\cite{QuarkDiquark}.

Let us mention that the tensor charge is not conserved and thus
depends on the scale $Q^2$. The $\chi$QSM scale is around
$Q^2_0=0.36$ GeV$^2$. The tensor charge at any scale $Q^2$ can be
obtained thanks to the evolution equation up to NLO \cite{Review}
\begin{equation}
\delta
q(Q^2)=\left(\frac{\alpha_s(Q^2)}{\alpha_s(Q_0^2)}\right)^{\frac{4}{27}}\left[1-\frac{337}{486\pi}(\alpha_s(Q_0^2)-\alpha_s(Q^2)\right]\delta
q(Q_0^2).
\end{equation}

Soffer \cite{Soffer} has proposed an inequality among the nucleon
twist 2 quark distributions $f_1,g_1,h_1$
\begin{equation}\label{Soff}
f_1+g_1\geq 2|h_1|
\end{equation}
In contrast to the well-known inequalities and positivity
constraints among distribution functions such as $f_1\geq |g_1|$
which are general properties of lepton-hadron scattering, derived
without reference to quarks, color or QCD, this Soffer inequality
needs a parton model to QCD to be derived \cite{Sofferineq}.
Unfortunately it turned out that it does not constrain the nucleon
tensor charge. However this inequality still has to be satisfied by
models that try to estimate quark distributions.

\section{Melosh rotation}\label{Section quatre}

In DIS one is probing the proton in IMF where the relativistic
many-body problem is suitably described. The usual light-cone
approach is to transform the instant quark states $\psi^i_\sigma$
into the light-cone quark states $\psi^i_{LC,\lambda}$, with
$i=1,2,3$. They are related by a general Melosh rotation
\cite{Melosh}
\begin{eqnarray}
\psi^i_{LC,+}&=&\frac{(m_q+z_i\uM)\psi^i_\uparrow+p_i^R\psi^i_\downarrow}{\sqrt{(m_q+z_i\uM)^2+\up_{i\perp}^2}},\\
\psi^i_{LC,-}&=&\frac{-p_i^L\psi^i_\uparrow+(m_q+z_i\uM)\psi^i_\downarrow}{\sqrt{(m_q+z_i\uM)^2+\up_{i\perp}^2}}
\end{eqnarray}
where $p_i^{R,L}=p_i^x\pm ip_i^y$ and $\uM$ is the invariant mass
$\uM^2=\sum_{i=1}^3(\up_i^2+m_q^2)/x_i$ with the constraints
$\sum_{i=1}^3 z_i=0$ and $\sum_{i=1}^3\up_{i\perp}=0$. The
zero-biding limit $z_i\uM\to p^+_i$ is not a justified approximation
for QCD bound states. This rotation mixes the helicity states due to
a nonzero transverse momentum $\up_{i\perp}$. The light-cone spinor
with helicity $+$ corresponds to \emph{total} angular momentum
projection $J_z=1/2$ and is thus constructed from a spin $\uparrow$
state with orbital angular momentum $L_z=0$ and a spin $\downarrow$
state with orbital angular momentum $L_z=1$ expressed by the factor
$p^R$. The light-cone spinor with helicity $-$ corresponds to
\emph{total} angular momentum projection $J_z=-1/2$ and is thus
constructed from a spin $\uparrow$ state with orbital angular
momentum $L_z=-1$ expressed by the factor $p^L$ and a spin
$\downarrow$ state with orbital angular momentum $L_z=0$.

The vector charge can be obtained in IMF by means of the \emph{plus}
component of the vector operator
\begin{equation}
q=\frac{1}{2}\langle
P,\frac{1}{2}|\bar\psi_{LC}\gamma^+\psi_{LC}|P,\frac{1}{2}\rangle.
\end{equation}
Using the Melosh rotation one can see that $q_{LC}$ and $q_{NR}$ are
related as follows
\begin{equation}
q_{LC}=\langle M_V\rangle q_{NR}
\end{equation}
where
\begin{equation}\label{Defq}
M_V=1.
\end{equation}

The axial charge can be obtained in IMF by means of the \emph{plus}
component of the axial operator
\begin{equation}
\Delta q=\frac{1}{2}\langle
P,\frac{1}{2}|\bar\psi_{LC}\gamma^+\gamma^5\psi_{LC}|P,\frac{1}{2}\rangle.
\end{equation}
Using the Melosh rotation one can see that $\Delta q_{LC}$ and
$\Delta q_{NR}$ are related as follows \cite{facteuraxial}
\begin{equation}
\Delta q_{LC}=\langle M_A\rangle\Delta q_{NR}
\end{equation}
where
\begin{equation}\label{DefMq}
M_A=\frac{(m_q+z_3\uM)^2-\up_{3\perp}^2}{(m_q+z_3\uM)^2+\up_{3\perp}^2}
\end{equation}
and $\langle M_A\rangle$ is its expectation value
\begin{equation}
\langle M\rangle=\int\ud^3p\,M|\Psi(p)|^2
\end{equation}
with $\Psi(p)$ a simple normalized momentum wave function. The
calculation with two different wave functions (harmonic oscillator
and power-law fall off) gave $\langle M_A\rangle=0.75$ \cite{Mq}.

The tensor charge can be obtained in IMF by means of the \emph{plus}
component of the tensor operator \cite{Schmidt}
\begin{equation}
\Delta q=\frac{1}{2}\langle
P,\frac{1}{2}|\bar\psi_{LC}\gamma^+\gamma^R\psi_{LC}|P,-\frac{1}{2}\rangle,
\end{equation}
where $\gamma^R=\gamma^1+i\gamma^2$. Using the Melosh rotation one
can see that $\delta q_{LC}$ and $\delta q_{NR}$ are related as
follows \cite{Schmidt}
\begin{equation}
\delta q_{LC}=\langle M_T\rangle\delta q_{NR}
\end{equation}
where
\begin{equation}\label{DefMqtilde}
M_T=\frac{(m_q+z_3\uM)^2}{(m_q+z_3\uM)^2+\up_{3\perp}^2}
\end{equation}
and $\langle M_T\rangle$ is its expectation value. In the
nonrelativistic limit $\up_\perp=0$ and thus $M_V=M_A=M_T=1$ as it
should be. One notices that relativistic effects $\up_\perp\neq 0$
reduce the values of $M_A$ and $M_T$. It is also interesting to
notice that one has
\begin{equation}
M_V+M_A=2M_T
\end{equation}
which saturates Soffer's inequality, see eq. (\ref{Soff}). Since
$\langle M_A\rangle=3/4$ one obtains $\langle M_T\rangle=7/8$ and
thus
\begin{equation}
\delta u=7/6,\qquad\delta d=-7/24,\qquad \delta s=0.
\end{equation}
From eqs. (\ref{DefMq}) and (\ref{DefMqtilde}) one would indeed
expect that
\begin{equation}
|\delta q|>|\Delta q|.
\end{equation}

In this approach the explicit valence wave function obtained
\cite{DiaPet} is
\begin{equation}\label{Discrete level IMF}
F^{j\sigma}_\textrm{lev}(z,\up_\perp)=\sqrt{\frac{\uM}{2\pi}}\left[\epsilon^{j\sigma}h(p)+(p_z\bold
1+i\up_\perp\times\tau_\perp)^\sigma_{\sigma'}\epsilon^{j\sigma'}\frac{j(p)}{|\up|}\right]_{p_z=z\uM-E_\textrm{lev}}
\end{equation}
to be compared with the Melosh rotated states
\begin{equation}
\psi^i_{LC,\lambda}=\frac{[(m+z_i\uM)\bold
1+i\un\cdot(\usigma\times\up_i)]^{\lambda'}_\lambda}{\sqrt{(m+z_i\uM)^2+\up^2_{i\perp}}}\,\psi^i_{NR,\lambda'}
\end{equation}
where $\un=(0,0,1)$. We have two functions $h(p)$ and $j(p)$
determined by the dynamics. The general form of a light-cone wave
function \cite{LFWF} should indeed contain two functions
\begin{equation}
\psi^\sigma_{\sigma_1}=\chi^\dagger_{\sigma_1}\left(f_1+\frac{i}{|\up|}\,\un\cdot(\usigma\times\up)f_2\right)\chi^\sigma.
\end{equation}
The additional $f_2$ term represents a separate dynamical
contribution to be contrasted with the purely kinematical
contribution of angular momentum from Melosh rotations. The $f_1$
term corresponds to states with $L_z=0$ and thus to $h$ and
$\frac{p_z}{|\up|}\,j$ while $f_2$ corresponds to states with
$L_z=\pm 1$ and thus to $\frac{p^{R,L}}{|\up|}\,j$. \newline In the
vector case, the one-quark line gives the contribution
\begin{equation}\label{Vectorform}
h^2(p)+2h(p)\,\frac{p_z}{|\up|}\,j(p)+j^2(p).
\end{equation}
In the axial case, the one-quark line gives the contribution
\begin{equation}\label{Axialform}
h^2(p)+2h(p)\,\frac{p_z}{|\up|}\,j(p)+\frac{2p_z^2-p^2}{p^2}\,j^2(p).
\end{equation}
In the tensor case, the one-quark line gives the contribution
\begin{equation}\label{Tensorform}
h^2(p)+2h(p)\,\frac{p_z}{|\up|}\,j(p)+\frac{p_z^2}{p^2}\,j^2(p).
\end{equation}
Clearly the connection with the Melosh rotation approach is achieved
by setting $h=0$ and $p_z=m+z\uM$. At the $3Q$ level the effect will
be similar, \emph{i.e.} all $SU(6)$ values are multiplied by a
common factor. It is however expected that this factor using the
general form of light cone wave function would be closer to 1 than
the one obtained by means of Melosh rotation.

\section{Currents, charges and matrix elements}\label{Section cinq}

A typical physical observable is the matrix element of some operator
(preferably written in terms of quark annihilation-creation
operators $a$, $b$, $a^\dag$, $b^\dag$) sandwiched between the
initial and final baryon wave functions. These wave functions are
superpositions of Fock states obtained by expanding the exponential
in eq. (\ref{Wavefunction}). One can reasonably expect that the Fock
states with the lowest number of quarks will give the main
contribution. If one uses the Drell frame $q^+=0$
\cite{Drell,Brodsky} where $q$ is the total momentum transfer then
the tensor $\bar \psi\gamma^+\gamma^\perp\psi$ current cannot create
nor annihilate any quark-antiquark pair. This is a big advantage of
the light-cone formulation since one needs to compute diagonal
transitions only, \emph{i.e.} $3Q$ into $3Q$, $5Q$ into $5Q$,
$\ldots$ and not $3Q$ into $5Q$ for example.

In the $3Q$ sector since all (valence) quarks are on the same
footing any contraction of creation-annihilation operators are
equivalent. One can use a diagram to represent these contractions.
The contractions without any current operator acting on a quark line
corresponds to the normalization of the state. We choose the
simplest one where all quarks with the same label are connected, see
Fig.\ref{Three-q normalization}.
\begin{figure}[h]\begin{center}\begin{minipage}[c]{8cm}\begin{center}\includegraphics[width=4cm]{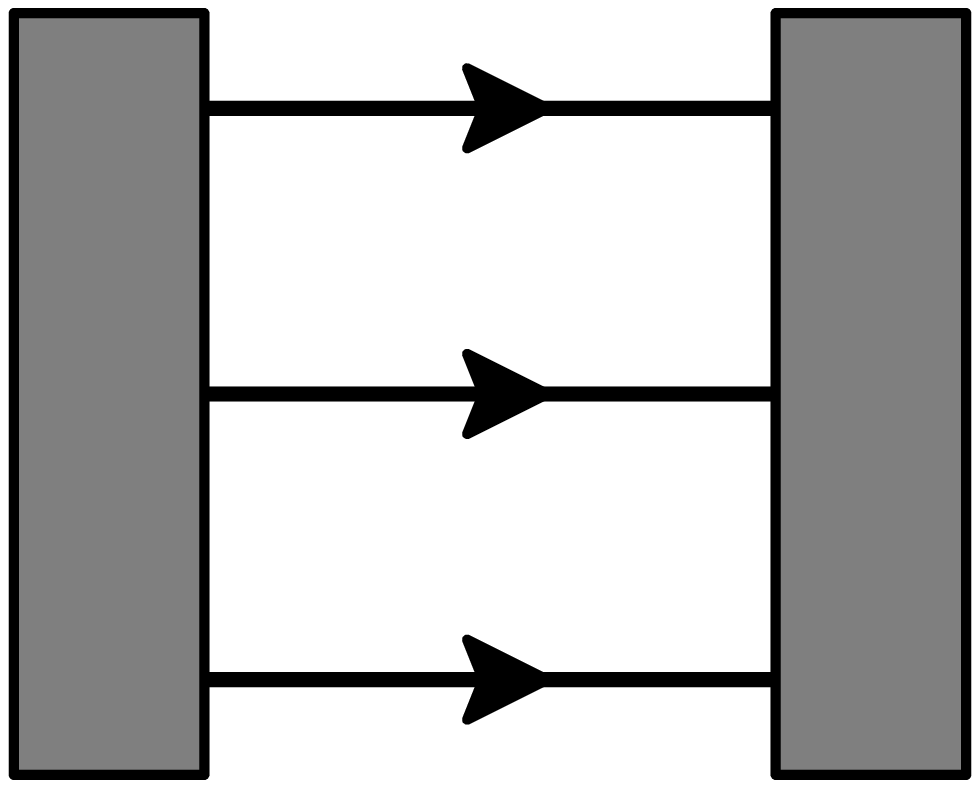}
\caption{\small{Schematic representation of the $3Q$ normalization.
Each quark line stands for the color, flavor and spin contractions
$\delta^{\alpha_i}_{\alpha'_i}\delta^{f_i}_{f'_i}\delta^{\sigma_i}_{\sigma'_i}
\int\ud
z'_i\,\ud^2\up'_{i\perp}\delta(z_i-z'_i)\delta^{(2)}(\up_{i\perp}-\up'_{i\perp})$.}}\label{Three-q
normalization}
\end{center}\end{minipage}\hspace{0.5cm}
\begin{minipage}[c]{8cm}\begin{center}\includegraphics[width=3.5cm]{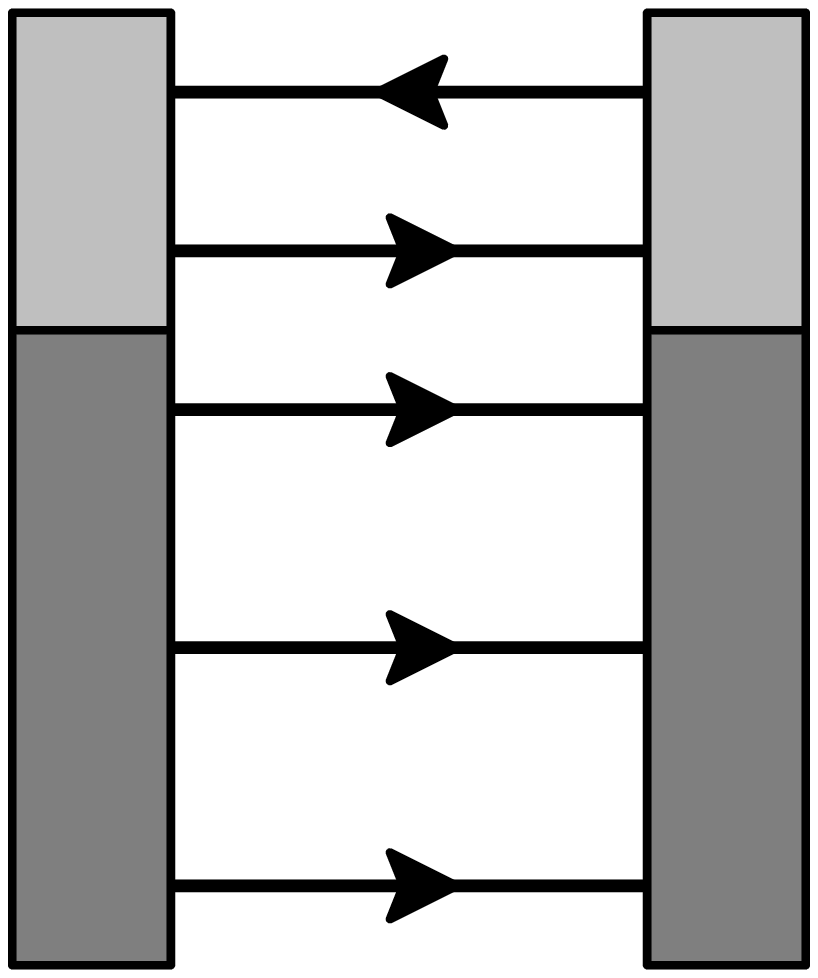}\hspace{1cm}\includegraphics[width=3.5cm]{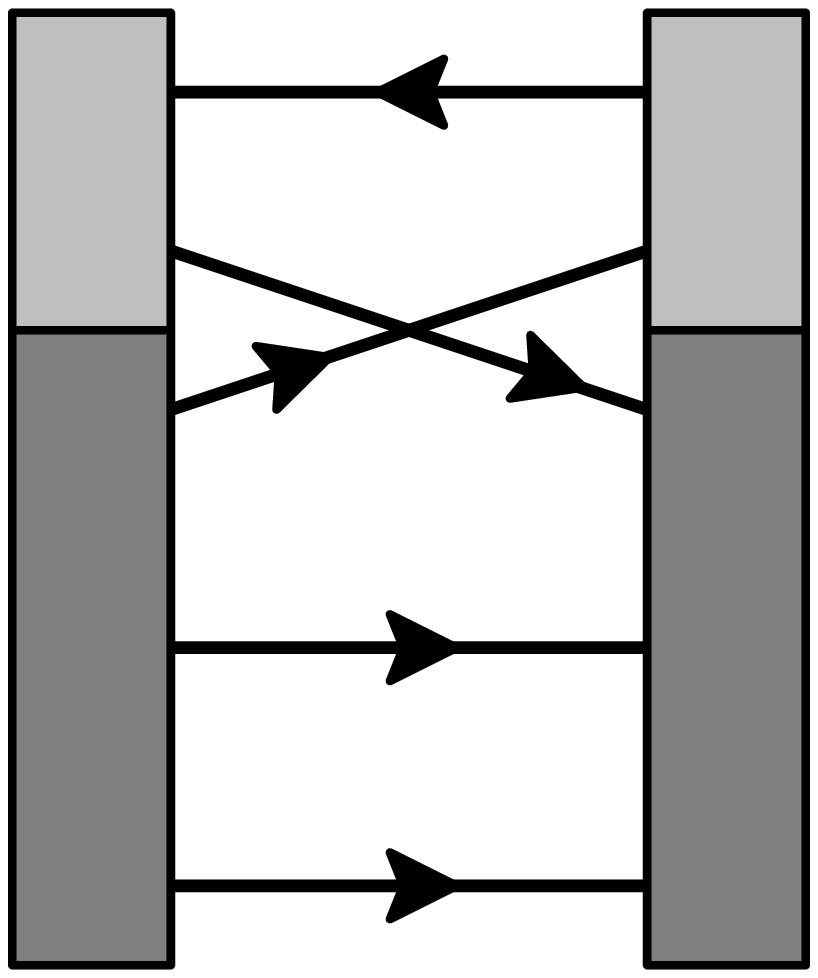}
\caption{\small{Schematic representation of the $5Q$ direct (left)
and exchange (right) contributions to the
normalization.}}\label{Five-q normalization}
\end{center}\end{minipage}\end{center}
\end{figure}

In the $5Q$ sector all contractions are equivalent to either the
so-called ``direct'' diagram or the ``exchange'' diagram, see
Fig.\ref{Five-q normalization}. In the direct diagram all quarks
with the same label are connected while in the exchange one a
valence quark is exchanged with the quark of the sea pair. It has
appeared in a previous work \cite{Moi} that exchange diagrams do not
contribute much and can thus be neglected. So in the $5Q$ sector we
used only the direct contribution in this paper.

\begin{figure}[h]\begin{center}\includegraphics[width=10cm]{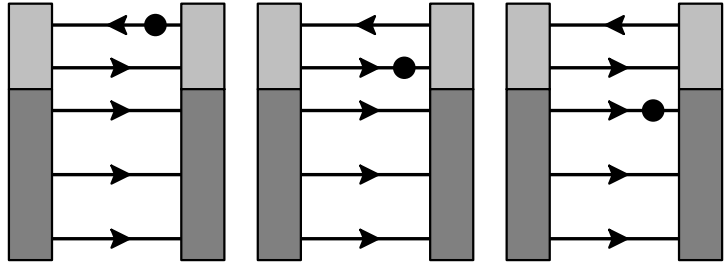}
\caption{\small{Schematic representation of the three types of $5Q$
contributions to the charges.}}\label{Direct charges}\end{center}
\end{figure}

The operator acts on each quark line. In the present approach it is
then easy to compute separately contribution coming from the valence
quarks, the sea quarks or antiquarks, see Fig.\ref{Direct charges}.
These diagrams represent some contraction of color, spin, isospin
and flavor indices. For example, the sum of the three diagrams in
the $5Q$ sector with the vector current $\bar\psi\gamma^+\psi$
acting on the quark lines represents the following expression
\begin{eqnarray}
V^{(5)}(1\to 2)&=&\frac{108}{2}\,\delta^k_lT(1)^{f_1f_2f_3f_4,j_5}_{j_1j_2j_3j_4,f_5,k}T(2)_{f_1f_2g_3g_4,l_5}^{l_1l_2l_3l_4,g_5,l}\int(\ud p_{1-5})\nonumber\\
&\times&F^{j_1\sigma_1}(p_1)F^{j_2\sigma_2}(p_2)F^{j_3\sigma_3}(p_3)W^{j_4\sigma_4}_{j_5\sigma_5}(p_4,p_5)F^\dag_{l_1\sigma_1}(p_1)F^\dag_{l_2\sigma_2}(p_2)F^\dag_{l_3\tau_3}(p_3)W_{c\,l_4\tau_4}^{l_5\tau_5}(p_4,p_5)\nonumber\\
&\times&\left[-\delta^{g_3}_{f_3}\delta^{g_4}_{f_4}\boldsymbol{
J^{f_5}_{g_5}}\delta^{\tau_3}_{\sigma_3}\delta^{\tau_4}_{\sigma_4}\boldsymbol{\delta_{\tau_5}^{\sigma_5}}+\delta^{g_3}_{f_3}\boldsymbol{
J^{g_4}_{f_4}}\delta^{f_5}_{g_5}\delta^{\tau_3}_{\sigma_3}\boldsymbol{\delta^{\tau_4}_{\sigma_4}}\delta_{\tau_5}^{\sigma_5}+3\boldsymbol{
J^{g_3}_{f_3}}\delta^{g_4}_{f_4}\delta^{f_5}_{g_5}\boldsymbol{\delta^{\tau_3}_{\sigma_3}}\delta^{\tau_4}_{\sigma_4}\delta_{\tau_5}^{\sigma_5}\right].\label{Direct}
\end{eqnarray}

\section{Scalar overlap integrals}\label{Section six}

The contractions in the previous section are easily performed by
\emph{Mathematica} over all flavor $(f,g)$, isospin $(j,l)$ and spin
$(\sigma,\tau)$ indices. One is then left with scalar integrals over
longitudinal $z$ and transverse $\up_\perp$ momenta of the quarks.
The integrals over relative transverse momenta in the $q\bar q$ pair
are generally UV divergent. We have chosen to use the Pauli-Villars
regularization with mass $M_\textrm{PV}=556.8$ MeV (this value being
chosen from the requirement that the pion decay constant $F_\pi=93$
MeV is reproduced from $M=345$ MeV).

For convenience we introduce the probability distribution
$\Phi^I(z,\uq_\perp)$ that three valence quarks leave the
longitudinal fraction $z=q_z/\uM$ and the transverse momentum
$\uq_\perp$ to the $q\bar q$ pair(s) with $I=V,T$ referring to the
vector or tensor case
\begin{equation}\label{Probability3q}
\Phi^I(z,\uq_\perp)=\int\ud
z_{1,2,3}\frac{\ud^2\up_{1,2,3\perp}}{(2\pi)^6}\,\delta(z+z_1+z_2+z_3-1)(2\pi)^2\delta^{(2)}(\uq_\perp+\up_{1\perp}+\up_{2\perp}+\up_{3\perp})D^I(p_1,p_2,p_3).
\end{equation}
The function $D^I(p_1,p_2,p_3)$ is given in terms of the upper and
lower valence wave functions $h(p)$ and $j(p)$ as follows
\begin{eqnarray}
D^V(p_1,p_2,p_3)&=&h^2(p_1)h^2(p_2)h^2(p_3)+6h^2(p_1)h^2(p_2)\left[h(p_3)\frac{p_{3z}}{|\up_3|}j(p_3)\right]+3h^2(p_1)h^2(p_2)j^2(p_3)\nonumber\\
&+&12h^2(p_1)\left[h(p_2)\frac{p_{2z}}{|\up_2|}j(p_2)\right]\left[h(p_3)\frac{p_{3z}}{|\up_3|}j(p_3)\right]+12h^2(p_1)\left[h(p_2)\frac{p_{2z}}{|\up_2|}j(p_2)\right]j^2(p_3)\nonumber\\
&+&8\left[h(p_1)\frac{p_{1z}}{|\up_1|}j(p_1)\right]\left[h(p_2)\frac{p_{2z}}{|\up_2|}j(p_2)\right]\left[h(p_3)\frac{p_{3z}}{|\up_3|}j(p_3)\right]+3h^2(p_1)j^2(p_2)j^2(p_3)\nonumber\\
&+&12\left[h(p_1)\frac{p_{1z}}{|\up_1|}j(p_1)\right]\left[h(p_2)\frac{p_{2z}}{|\up_2|}j(p_2)\right]j^2(p_3)+6\left[h(p_1)\frac{p_{1z}}{|\up_1|}j(p_1)\right]j^2(p_2)j^2(p_3)\nonumber\\
&+&j^2(p_1)j^2(p_2)j^2(p_3)\label{Phi}\\\nonumber\\
D^T(p_1,p_2,p_3)&=&h^2(p_1)h^2(p_2)h^2(p_3)+6h^2(p_1)h^2(p_2)\left[h(p_3)\frac{p_{3z}}{|\up_3|}j(p_3)\right]+h^2(p_1)h^2(p_2)\frac{p_{3z}^2+2p_3^2}{p_3^2}j^2(p_3)\nonumber\\
&+&12h^2(p_1)\left[h(p_2)\frac{p_{2z}}{|\up_2|}j(p_2)\right]\left[h(p_3)\frac{p_{3z}}{|\up_3|}j(p_3)\right]+4h^2(p_1)\left[h(p_2)\frac{p_{2z}}{|\up_2|}j(p_2)\right]\frac{p_{3z}^2+2p_3^2}{p_3^2}j^2(p_3)\nonumber\\
&+&8\left[h(p_1)\frac{p_{1z}}{|\up_1|}j(p_1)\right]\left[h(p_2)\frac{p_{2z}}{|\up_2|}j(p_2)\right]\left[h(p_3)\frac{p_{3z}}{|\up_3|}j(p_3)\right]+h^2(p_1)j^2(p_2)\frac{2p_{3z}^2+rp_3^2}{p_3^2}j^2(p_3)\nonumber\\
&+&4\left[h(p_1)\frac{p_{1z}}{|\up_1|}j(p_1)\right]\left[h(p_2)\frac{p_{2z}}{|\up_2|}j(p_2)\right]\frac{p_{3z}^2+2p_3^2}{p_3^2}j^2(p_3)+2\left[h(p_1)\frac{p_{1z}}{|\up_1|}j(p_1)\right]j^2(p_2)\frac{2p_{3z}^2+p_3^2}{p_3^2}j^2(p_3)\nonumber\\
&+&j^2(p_1)j^2(p_2)\frac{p_{3z}^2}{p_3^2}j^2(p_3).\label{Psi}
\end{eqnarray}
In the nonrelativistic limit one has $j(p)=0$ and thus
$D^V(p_1,p_2,p_3)=D^T(p_1,p_2,p_3)$. The expression for the axial
case can be found in \cite{Moi3}.

\subsection{$3Q$ scalar integrals}

In the $3Q$ sector there is no quark-antiquark pair. There are then
two integrals only, $\Phi^V(0,0)$ and $\Phi^T(0,0)$. Let us remind
that in this sector spin-flavor wave functions obtained by the
projection technique are equivalent to those given by $SU(6)$
symmetry. One then naturally obtains the same results than given by
$SU(6)$ excepted that tensor quantities are multiplied by the factor
$\Phi^T(0,0)/\Phi^V(0,0)$. As discussed earlier this is similar (but
not exactly the same) to approaches using Melosh rotations.

\subsection{$5Q$ scalar integrals}

In the $5Q$ sector there is one quark-antiquark pair and only six
integrals are needed. These integrals can be written in the general
form
\begin{equation}
K^I_J=\frac{M^2}{2\pi}\int \frac{\ud^3
\uq}{(2\pi)^3}\,\Phi^I\left(\frac{q_z}{\uM},\uq_\perp\right)\theta(q_z)\,q_z\,G_J(q_z,\uq_\perp)
\end{equation}
where $G_J$ is a quark-antiquark probability distribution and
$J=\pi\pi,33,\sigma\sigma$. These distributions are obtained by
contracting two quark-antiquark wave functions $W$, see eq.
(\ref{Pair wavefunction}) and regularized by means of Pauli-Villars
procedure
\begin{eqnarray}
G_{\pi\pi}(q_z,\uq_\perp)&=&\Pi^2(\uq)\int^1_0\ud
y\int\frac{\ud^2\uQcal_\perp}{(2\pi)^2}\frac{\uQcal^2_\perp+M^2}{(\uQcal^2_\perp+M^2+y(1-y)\uq^2)^2}-(M\to
M_\textrm{PV}),\label{Direct1}
\\
G_{33}(q_z,\uq_\perp)&=&\frac{q_z^2}{\uq^2}\,G_{\pi\pi}(q_z,\uq_\perp),\label{Direct2}
\\
G_{\sigma\sigma}(q_z,\uq_\perp)&=&\Sigma^2(\uq)\int^1_0\ud
y\int\frac{\ud^2\uQcal_\perp}{(2\pi)^2}\frac{\uQcal^2_\perp+M^2(2y-1)^2}{(\uQcal^2_\perp+M^2+y(1-y)\uq^2)^2}-(M\to
M_\textrm{PV}),\label{Direct3}
\end{eqnarray}
where $q_z=z\uM=(z_4+z_5)\uM$ and
$\uq_\perp=\up_{4\perp}+\up_{5\perp}$. There are three integrals in
the vector case $K^V_{\pi\pi},K^V_{33},K^V_{\sigma\sigma}$ and three
in the tensor one $K^T_{\pi\pi},K^T_{33},K^T_{\sigma\sigma}$. Sea
quarks and antiquarks do not contribute to the tensor charge since
the the tensor operator is chiral-odd. In this approach it is
reflected by the fact that the contraction of two quark-antiquark
wave functions $W$ with the tensor operator leaves only vanishing
scalar overlap integrals $\int\ud^2\uQcal_\perp\,\uQcal_x$ or
$\int\ud^2\uQcal_\perp\,\uQcal_y$.

Even though sea quarks and antiquarks do not contribute to the
tensor charge it is not sufficient to restrict the computation to
the $3Q$ sector where only valence quarks appear. Higher Fock states
change the composition of the valence sector as shown by eq.
(\ref{hidden}). So hidden flavors can access to the valence level
and thus contribute to tensor charge of the baryon. In other words,
even though only valence quarks contribute $SU(6)$ relations are
broken due to relativistic effects (additional quark-antiquark
pairs).

\section{Results}\label{Section sept}

\subsection{Combinatoric results}

In this work we have studied tensor charges in flavor $SU(3)$
symmetry. Even though this symmetry is broken in nature, this gives
quite a good estimation. The interesting thing is that this symmetry
relates tensor charges within each multiplet. Indeed all particles
in a given representation are on the same footing and are related
through pure flavor $SU(3)$ transformations. One can find the way to
relate tensors charges of different members of the same multiplet in
\cite{Moi3}.

The octet, decuplet and antidecuplet normalizations in the $3Q$ and
$5Q$ sectors are given by the following linear combination
\begin{eqnarray}
\uN^{(3)}(B_{\bf 8})&=&9\,\Phi^V(0,0),\\
\uN^{(5)}(B_{\bf
8})&=&\frac{18}{5}\left(11K^V_{\pi\pi}+23K^V_{\sigma\sigma}\right),\\
\nonumber\\
\uN^{(3)}_{3/2}(B_{\bf 10})&=&\uN^{(3)}_{1/2}(B_{\bf 10})=\frac{18}{5}\,\Phi^V(0,0),\\
\uN^{(5)}_{3/2}(B_{\bf
10})&=&\frac{9}{5}\left(15K^V_{\pi\pi}-6K^V_{33}+17K^V_{\sigma\sigma}\right),\\
\uN^{(5)}_{1/2}(B_{\bf
10})&=&\frac{9}{5}\left(11K^V_{\pi\pi}+6K^V_{33}+17K^V_{\sigma\sigma}\right),\\
\nonumber\\
\uN^{(5)}(B_{\bf\overline{10}})&=&\frac{36}{5}\left(K^V_{\pi\pi}+K^V_{\sigma\sigma}\right)
\end{eqnarray}
where the subscript $3/2,1/2$ refers to the value of third component
of the baryon spin $J_z$.

Here are the proton tensor charges
\begin{eqnarray}
T^{(3)}_u(p)&=&12\Phi^T(0,0),\\
T^{(3)}_d(p)&=&-3\Phi^T(0,0),\\
T^{(3)}_s(p)&=&0,\\\nonumber\\
T^{(5)}_u(p)&=&\frac{18}{25}\left(48 K^T_{\pi\pi}-7K^T_{33}+151K^T_{\sigma\sigma}\right),\\
T^{(5)}_d(p)&=&\frac{-12}{25}\left(24 K^T_{\pi\pi}+19K^T_{33}+53K^T_{\sigma\sigma}\right),\\
T^{(5)}_s(p)&=&\frac{-12}{25}\left(3
K^T_{\pi\pi}+8K^T_{33}+K^T_{\sigma\sigma}\right).
\end{eqnarray}

Here are the $\Delta^{++}$ tensor charges
\begin{eqnarray}
T^{(3)}_{u,3/2}(\Delta^{++})&=&\frac{54}{5}\Phi^T(0,0),\\
T^{(3)}_{d,3/2}(\Delta^{++})&=&T^{(3)}_{s,3/2}(B_{\bf 10})=0,\\
T^{(3)}_{u,1/2}(\Delta^{++})&=&\frac{18}{5}\Phi^T(0,0),\\
T^{(3)}_{d,1/2}(\Delta^{++})&=&T^{(3)}_{s,1/2}(\Delta^{++})=0,\\\nonumber\\
T^{(5)}_{u,3/2}(\Delta^{++})&=&\frac{9}{10}\left(56 K^T_{\pi\pi}-17K^T_{33}+101K^T_{\sigma\sigma}\right),\\
T^{(5)}_{d,3/2}(\Delta^{++})&=&T^{(5)}_{s,3/2}(B_{\bf
10})=\frac{-9}{20}\left(8
K^T_{\pi\pi}+13K^T_{33}-K^T_{\sigma\sigma}\right),\\
T^{(5)}_{u,1/2}(\Delta^{++})&=&\frac{3}{10}\left(42 K^T_{\pi\pi}+25K^T_{33}+101K^T_{\sigma\sigma}\right),\\
T^{(5)}_{d,1/2}(\Delta^{++})&=&T^{(5)}_{s,1/2}(\Delta^{++})=\frac{-3}{20}\left(6
K^T_{\pi\pi}+19K^T_{33}-K^T_{\sigma\sigma}\right).
\end{eqnarray}

Here are the $\Theta^+$ tensor charges
\begin{eqnarray}
T^{(5)}_u(\Theta^+)&=&T^{(5)}_d(\Theta^+)=\frac{-18}{5}\left(K^T_{33}-K^T_{\sigma\sigma}\right),\\
T^{(5)}_s(\Theta^+)&=&0.
\end{eqnarray}
In the $5Q$ sector of $\Theta^+$ pentaquark the strange flavor
appears only as an antiquark as one can see from its minimal quark
content $uudd\bar s$. That's the reason why we have found no strange
contribution. But if at least the $7Q$ sector was considered we
would have obtained  a nonzero contribution due flavor components
like $|uus(d\bar s)(d\bar s)\rangle$, $|uds(u\bar s)(d\bar
s)\rangle$ and $|dds(u\bar s)(u\bar s)\rangle$.

\subsection{Numerical results}

In the evaluation of the scalar integrals we have used the
constituent quark mass $M=345$ MeV, the Pauli-Villars mass
$M_\textrm{PV}=556.8$ MeV for the regularization of
(\ref{Direct1})-(\ref{Direct3}) and the baryon mass $\uM=1207$ MeV
as it follows for the ``classical'' mass in the mean field
approximation \cite{Approximation}. Choosing $\Phi^V(0,0)=1$ we have
obtained in the $3Q$ sector
\begin{equation}
\Phi^T(0,0)=0.9306
\end{equation}
and in the $5Q$ sector
\begin{eqnarray}
&K^V_{\pi\pi}=0.0365,\qquad K^V_{33}=0.0197,\qquad
K^V_{\sigma\sigma}=0.0140,&\\\nonumber\\
&K^T_{\pi\pi}=0.0333,\qquad K^T_{33}=0.0180,\qquad
K^T_{\sigma\sigma}=0.0126.&
\end{eqnarray}
This has to be compared with the results in the axial case
\cite{Moi}
\begin{eqnarray}
&\Phi^A(0,0)=0.8612&\\\nonumber\\
&K^A_{\pi\pi}=0.0300,\qquad K^A_{33}=0.0163,\qquad
K^A_{\sigma\sigma}=0.0112.&
\end{eqnarray}
As expected from (\ref{Vectorform}), (\ref{Axialform}) and
(\ref{Tensorform}) we have the following pattern for the integrals
$|V|>|T|>|A|$.

\subsection{Discussion}

We collect in Tables \ref{proton}, \ref{delta} and \ref{theta} our
results concerning the tensor charges at the model scale
$Q_0^2=0.36$ GeV$^2$.

\begin{table}[h!]\begin{center}\caption{\small{Our proton tensor charges and their isovector and isoscalar combinations.\newline}}
\begin{tabular}{c|ccc|cc}
\hline\hline $p^+$&$\delta u$&$\delta d$&$\delta
s$&$g_T^{(3)}$&$g_T^{(0)}$\rule{0pt}{3ex}\\\hline \rule{0pt}{3ex}
$3Q$&1.241&-0.310&0&1.551&0.931\\\rule{0pt}{3ex}
$3Q+5Q$&1.172&-0.315&-0.011&1.487&0.846\rule[-2ex]{0pt}{5ex}\\
\hline\hline
\end{tabular}\label{proton}\end{center}
\end{table}
\begin{table}[h!]\begin{center}\caption{\small{Our $\Delta^{++}$ tensor charges and their isovector and isoscalar combinations.\newline}}
\begin{tabular}{c|ccc|cc}
\hline\hline $\Delta^{++}_{3/2}$&$\delta u$&$\delta d$&$\delta
s$&$g_T^{(3)}$&$g_T^{(0)}$\rule{0pt}{3ex}\\\hline \rule{0pt}{3ex}
$3Q$&2.792&0&0&2.792&2.792\\\rule{0pt}{3ex}
$3Q+5Q$&2.624&-0.046&-0.046&2.670&2.532\rule[-2ex]{0pt}{5ex}\\\hline\rule{0pt}{3ex}
$\Delta^{++}_{1/2}$&$\delta u$&$\delta d$&$\delta
s$&$g_T^{(3)}$&$g_T^{(0)}$\rule{0pt}{3ex}\\\hline \rule{0pt}{3ex}
$3Q$&0.931&0&0&0.931&0.931\\\rule{0pt}{3ex}
$3Q+5Q$&0.863&-0.016&-0.016&0.879&0.831\rule[-2ex]{0pt}{5ex}\\
\hline\hline
\end{tabular}\label{delta}\end{center}
\end{table}
\begin{table}[h!]\begin{center}\caption{\small{Our $\Theta^+$ tensor charges and their isovector and isoscalar combinations.\newline}}
\begin{tabular}{c|ccc|cc}
\hline\hline $\Theta^+$&$\delta u$&$\delta d$&$\delta
s$&$g_T^{(3)}$&$g_T^{(0)}$\rule{0pt}{3ex}\\\hline \rule{0pt}{3ex}
$3Q+5Q$&-0.053&-0.053&0&0&-0.107\rule[-2ex]{0pt}{5ex}\\
\hline\hline
\end{tabular}\label{theta}\end{center}
\end{table}
Like all other models for the proton $\delta u$ and $\delta d$ are
not small and have a magnitude similar to $\Delta u$ and $\Delta d$.
One can also check that Soffer's inequality (\ref{Soff}) is
satisfied for explicit flavors. However hidden flavors, \emph{i.e.}
$s$ in proton and $d,s$ in $\Delta^{++}$, violate the inequality.

Up to now only one experimental extraction of transversity
distributions has been achieved \cite{Exptensor}. The authors did
not give explicit values for tensor charges. They have however been
estimated to $\delta u=0.46^{+0.36}_{-0.28}$ and $\delta
d=-0.19^{+0.30}_{-0.23}$ in \cite{QuarkDiquark} at the scale
$Q^2=0.4$ GeV$^2$. These values are unexpectedly small compared to
models predictions. Further experimental results are then highly
desired to either confirm or infirm the smallness of tensor charges.

\section{Conclusion}

We have used Chiral Quark Soliton Model ($\chi$QSM) formulated in
the Infinite Momentum Frame (IMF) up to $5Q$ Fock component to
investigate octet, decuplet and antidecuplet tensor charges. We have
obtained $\delta u=1.172$ and $\delta d=-0.315$ at $Q_0^2=0.36$
GeV$^2$ for the proton which are in the range of prediction of the
other models.

We have also discussed the Melosh rotations involved in usual
light-cone approach compared with our approach. Melosh rotation
introduces somewhat artificially angular momentum whose origin is
purely kinematical. A general light-cone wave function should in
fact contain a dynamical term like in the approach used in this
paper.

Usual light-cone models consider only the $3Q$ sector and thus
cannot estimate the strange tensor charge $\delta s$. Even though
sea quarks and antiquarks do not contribute to tensor charges one
can obtain a nonzero $\delta s$ because the $5Q$ component of the
nucleon allows strange quarks to access to the valence level. Our
result is $\delta s=-0.011$ and thus a negative transverse
polarisation of strange quarks.

\subsection*{Acknowledgements}

The author is grateful to RUB TP2 for its kind hospitality and to M.
Polyakov for his careful reading and comments. The author is also
indebted to J. Cugnon whose absence would not have permitted the
present work to be done. This work has been supported by the
National Funds of Scientific Research, Belgium.

\end{document}